\documentclass[aps,amssymb,amsmath,twocolumn,showpacs,showkeys]{revtex4}
\usepackage{graphicx}
\newcommand{\ket}[1]{\left|{#1}\right\rangle}
\newcommand{\bra}[1]{\left\langle{#1}\right|}
\newcommand{\modu}[1]{\left|{#1}\right|}
\newcommand{\expect}[3]{\left\langle{#1}\right|{#2}\left|{#3}\right\rangle}
\newcommand{\aver}[1]{\left\langle{#1}\right\rangle}
\newcommand{\added}[2]{\left|{#1},{#2}\right\rangle}
\newcommand{\inner}[2]{\left\langle{#1}|{#2}\right\rangle}
\newcommand{\beq}{\begin{equation}}
\newcommand{\eeq}{\end{equation}}
\begin{document}
\title{Non-classical properties of quantum wave packets propagating
in a Kerr-like medium}
\author{C. Sudheesh, S. Lakshmibala, and V. Balakrishnan}
\email{sudheesh,slbala,vbalki@physics.iitm.ac.in}
\affiliation{
Department of Physics, Indian Institute of Technology Madras,
Chennai 600 036, India}
\date{\today}
\begin{abstract}
We investigate  non-classical effects such as fractional 
revivals, squeezing and 
higher-order squeezing of 
photon-added coherent states propagating through a Kerr-like medium.
The Wigner functions corresponding to these states 
at the instants of fractional revivals are obtained,  
and the extent of non-classicality quantified. 
\end{abstract} 

\maketitle
\section{Introduction}
Non-classical effects such as revivals 
and squeezing exhibited by quantum wave packets while 
propagating
through different media are of great interest in the context of   
quantum information processing with continuous variables\cite{brau}. 
Both the nature of the medium in which the state propagates
and the precise initial state considered play a crucial role in 
determining
the subsequent dynamics of the state. 
With the experimental realization 
of several non-classical states in quantum
optics\cite{dodo}, it has become important  to
examine more closely their
dynamical properties.
Detailed investigations have 
been carried out, for instance, on the revival phenomena displayed by an 
initial coherent 
state evolving under different Hamiltonians\cite{robi}. In particular,  
the non-classical features displayed by a wave packet which is initially 
a coherent state while it 
propagates through a nonlinear, Kerr-like medium have been examined. 
Its squeezing and
higher-order squeezing properties have been discussed\cite{du}, 
and signatures of wave packet revivals 
and fractional revivals have been shown to be captured in the expectation 
values of appropriate observables \cite{sudh}.

In this paper, we examine the  effects of the {\it departure} 
from coherence of 
the initial state on the subsequent non-classical features it displays
while evolving in a Kerr-like medium. In order to be able to 
compare with exactitude these effects  
with those that arise in the dynamics of an initial coherent state 
(CS), it is essential to 
consider such  initial states as display a precisely 
quantifiable departure 
from coherence. 
The photon-added coherent states (PACS) are promising candidates  
for this purpose  
\cite{agar}.
We examine a spectrum of 
non-classical phenomena  such as revivals, fractional revivals, squeezing 
and higher-order squeezing displayed by an initial PACS as it propagates 
in a Kerr-like medium, and identify the effects arising from the  
departure from perfect coherence of the initial state. 

The plan of the paper is as follows: In the next section, we  
review briefly the salient features of revivals and fractional 
revivals of 
a wave packet, and identify quantifiers which 
provide signatures to distinguish between fractional revivals of 
initial wave packets that are ideal CS from those that are PACS.
In Section III, we examine 
the dependence of squeezing and higher-order squeezing effects 
on the extent of coherence of the initial 
wave packet. In the final section, we estimate the degree of non-classicality 
displayed by a CS and by a PACS at specific instants while evolving in 
a Kerr-like medium, by quantifying the extent of 
negativity of the corresponding Wigner 
functions.

\section{Wave Packet Revivals}

A wave packet propagating through a nonlinear medium may display 
revivals and fractional revivals at specific instants of time.
In the one-dimensional case, if the initial wave packet
$\ket{\psi(0)}$ is a superposition of a sufficient number
of basis states, it is expected to revive  at periodic intervals of
time, though these intervals 
could be quite long in practice\cite{guts}. 
A full  revival at time $T$ is signaled
by the fact that the autocorrelation
function $\modu{\inner{\psi(0)}{\psi(T)}}^2$ returns to its initial
value of unity. 

In a generic nonlinear 
medium, for wave packets which are peaked 
sufficiently sharply in energy about 
some eigenvalue labeled by $n_0$, 
we may expand the energy spectrum $E_n$ in a 
Taylor series about $E_{n_0}\,$, and retain only terms 
up to the second order in $(n - n_0)\,$. 
It can then be shown\cite{aver} that fractional 
revivals  of the wave packet occur at certain instants of time between 
successive revivals, when the wave packet 
splits into a superposition of a finite number of 
spatially separated sub-packets, each of 
which closely resembles the original one. This non-classical 
phenomenon arises due to very specific quantum interference properties 
between the basis states comprising the original wave packet\cite{tara}. 
The dynamics of a 
wave packet evolving in a Kerr-like medium is governed by the model
Hamiltonian 
\begin{equation}
H = \hbar \chi a^{\dagger 2}a^2 = \hbar \chi N(N-1),
\label{hamil}
\end{equation}
where $N = a^\dagger a$ and $a\,,\,a^\dagger$ are photon annihilation 
and creation operators, so that $[a , a^\dagger] = 1$. The positive 
constant $\chi$ represents the susceptibility of the nonlinear 
medium. (The same Hamiltonian is applicable to a Bose-Einstein 
condensate propagating 
in a three-dimensional optical lattice. In this case, $a$ and $a^\dagger$  
are atom annihilation and creation operators, and $\chi$ 
characterizes the energy needed to overcome the inter-atomic repulsion.)

For ready reference, we first state the results\cite{sudh} 
for an initial
coherent state $\ket{\psi(0)} = \ket{\alpha}$,
where $a \ket{\alpha} = \alpha
\ket{\alpha}$ and the complex number 
$\alpha \equiv (x_o+ip_0)/\sqrt{2}$ labels the CS.
Such an initial state can be shown to revive 
periodically, with a revival time 
\begin{equation}
T_{\rm rev} = \pi/\chi. 
\label{trev}
\end{equation}
Between $t = 0$ and $t = T_{\rm rev}\,$, fractional revivals 
occur at times $t = \pi j/(k\chi)$, where $k = 2,\,3,\ldots$ and $j =1,\,2,
\ldots,\, (k-1)$ for a given value of $k$. At time 
$(j/k)T_{\rm rev}$, the initial
Gaussian wave packet  
splits into $k$ spatially distributed sub-packets that are similar to
itself. 
Full revivals of the wave packet can be discerned by observing 
the time evolution of the expectation values (first moments) 
of the operators $x\equiv (a+a^\dagger)/\sqrt{2}$ and  $p \equiv
(a-a^\dagger)/(i\sqrt{2})$, while
distinctive signatures of the $k$-sub-packet fractional revivals 
manifest themselves in the temporal behavior  
of the $k^{\rm th}$ moments of $x$ and $p$. 
The time dependence of all moments of $x$ and $p$ can be obtained from
the general result
\begin{eqnarray}
\aver{a^{\dagger r}a^{r+s}}&\equiv&
\expect{\psi(t)}{a^{\dagger 
r}\,a^{r+s}}{\psi(t)}\nonumber\\
&=& \alpha^{s}\,\nu^{r}
e^{-\nu\,(1-\cos \,2 s \chi t)}\nonumber\\
&\times&e^{
-i \chi [s(s-1)
+ 2rs]\,t - i\nu \,\sin\,2s \chi t},
\label{csexpect}
\end{eqnarray}
where $r$ and $s$ are non-negative integers, and 
$\nu \equiv \modu{\alpha}^2$ is the mean 
number of photons in the CS. Writing 
$\expect{\psi(t)}{x}{\psi(t)}=\aver{x(t)}$, etc., 
it is evident that 
$\aver{x(0)} = x_0$ and $\aver{p(0)}= p_0\,$. We find
\noindent
\begin{eqnarray}
\aver{x(t)}&=&
e^{-\nu\,(1-\cos 2\chi t)}\,
\big[ x_0 \,\cos \,(\nu \sin 2\chi t)\nonumber\\
&+&p_0\, 
\sin \,(\nu \sin 2\chi t)\big],
\label{xt}
\end{eqnarray}
with a similar expression for $\aver{p(t)}$. 
Owing to the exponential factor on the RHS we find that, 
for sufficiently large values of $\nu$, 
$\aver{x(t)}$ remains essentially equal to 
$\aver{x(0)}$ for much of the time, except for significant and rapid
variation near instants of full revival. This behavior is shared by 
$\aver{p(t)}$. A sudden 
change in $\aver{x(t)}$ or $\aver{p(t)}$ 
thus signals the occurrence 
of a revival of the initial wave packet. 
Similarly, writing  
$\expect{\psi(t)}{x^2}{\psi(t)}=
 \aver{x^2(t)}$, we can show that
\begin{eqnarray}
2 \aver{x^2(t)}&=& 1 
+ x_{0}^{2} + p_{0}^{2}  + e^{-\nu\,(1-\cos \,4\chi t)} \nonumber\\
&\times&\,\big[ (x_{0}^{2} - p_{0}^{2})\,\cos\,  
(2\chi t + \nu \sin \,4\chi t) \nonumber\\ 
&+& 2 x_{0} p_{0} \,\sin\, (2\chi t + \nu\sin \,4\chi t)\big].
\label{xsqrdt}
\end{eqnarray}
Once again, for large values of $\nu$, $\aver{x^2(t)}$ is more or less
static around the value $\aver{x^2(0)}$ except at times close to 
$\pi/{(2\chi)}= \frac{1}{2}T_{\rm rev}$, when 
the 2-sub-packet fractional revival occurs. 
Similar behavior is displayed by 
$\aver{p^2(t)}$. An explicit calculation reveals that, in general,
 the 
signature 
of a $k$-sub-packet fractional revival is mirrored in the dynamics of 
$\aver{x^k(t)}$ (equivalently, of $\aver{p^k(t)}$). 
It follows that the plot of $\varDelta x$ (the
standard deviation in $x$)  versus $\varDelta p$ (the standard deviation in
$p$) captures the occurrence of a 2-sub-packet fractional revival. 
Likewise, the
square of the skewness in $x$ (or $p$)  
mirrors the occurrence of a 3-sub-packet
fractional revival, while the temporal behaviour of the kurtosis of $x$ (or
$p$) enables us to detect the appearance of a 4-sub-packet fractional
revival.

Next, we turn to the case in which the initial state is 
a PACS rather than a (perfectly coherent) CS.  
The normalized $m$-photon-added coherent state 
$\added{\alpha}{m} \,(m=1,\, 2, 
\ldots)$ is defined as 
\begin{equation}
\added{\alpha}{m}
=\frac{(a^\dagger)^m\ket{\alpha}}{\sqrt{\expect{\alpha}{a^m 
\,a^{\dagger m}}{\alpha}}}
=\frac{(a^\dagger)^m\ket{\alpha}}{\sqrt{
m!\,L_{m}(-\nu)}},
\label{photonadded}
\end{equation} 
where $L_m$ is the Laguerre polynomial of order $m$. The PACS just
defined is also a {\it nonlinear} coherent state in the sense that
it is an eigenstate of a nonlinear
annihilation operator, namely, 
\begin{equation}
\left(1- \frac{m}{1+a^\dagger a}\right)a\added{\alpha}{m}= \alpha 
\added{\alpha}{m}.
\label{nonlinearcs}
\end{equation}
A PACS of this sort 
can be produced in laser-atom 
interactions under appropriate conditions\cite{agar}. 
Clearly, with increasing $m$ (the number of photons `added' to the ideal 
CS $\ket{\alpha}$), there is increasing departure from perfect 
coherence. It is straightforward to see that, when an initial 
state $\added{\alpha}{m}$ evolves 
under the Hamiltonian in Eq. (\ref{hamil}), 
revivals and fractional revivals of the 
wave packet occur at the same instants as in the case of a coherent
initial state, and that their
appearance is mirrored in the time evolution of the expectation 
values of the different moments of the quadratures 
$x$ and $p$, as before. However, the 
actual dynamics of these expectation values is much more complex, even 
for small values of $m$: even a small departure 
from  Poisson number statistics in the initial state 
leads to significant changes in 
the evolution of the wave packet.

To illustrate this explicitly, we find first
the counterpart 
of Eq. (\ref{csexpect}) for the general initial state 
$ \ket{\psi(0)} = \added{\alpha}{m}$. The result can be written in
closed form as  
\begin{eqnarray}
\lefteqn{\aver{a^{\dagger r}\,a^{r+s}}_m
=\alpha^s e^{-\nu+\nu \cos\,2s\chi t 
-i\chi( s-1+2m)st
-i\nu \sin\,2s\chi t}}\nonumber\\
&&\times 
\sum_{n=0}^{r}\,\binom{r}{n}\,\frac{m!\,(\nu e^{-2is\chi t})^n}{(m-r+n)!}
\frac{L_m^{s+n}(-\nu e^{-2is\chi t})}{L_m(-\nu)}, 
\label{pacsexpect}
\end{eqnarray}
where $L_m^{n}$ denotes the associated Laguerre 
polynomial. Setting $m=0$, we recover  Eq. (\ref{csexpect}) for an initial 
state $\ket{\alpha}$.  
For an initial state $\ket{\alpha, 1}$, 
corresponding to minimal departure from coherence in this class of PACS, 
we set $r = 0$, $s = 1$ 
and $m=1$ in Eq. (\ref{pacsexpect}) to obtain the expectation value of 
$a(t)$, and hence the expectation value of $x(t)$, denoted by 
$\aver{x(t)}_1$. We get
\begin{equation}
\aver{x(t)}_1 = e^{-\nu(1-\cos 2\chi t)}
[x_0\,{\rm Re}\,z_1(t) + p_0\,{\rm Im}\, z_1(t)],
\label{xt1}
\end{equation}
where
\begin{equation}
z_1 =\left(\frac{2+\nu\, e^{2i\chi t}}
{1 + \nu}\right) 
e^{i(2\chi t+\nu\,\sin \,2\chi t)}.
\label{z1}
\end{equation}
Similarly, for the initial state $\ket{\alpha, 2}$ 
we find
\begin{equation}
\aver{x(t)}_2 = e^{-\nu(1-\cos 2\chi t)}
[x_0\,{\rm Re}\,z_2(t) + p_0\,{\rm Im}\, z_2(t)],
\label{xt2}
\end{equation}
where
\begin{equation}
z_2 =\left(\frac{6+6\nu \,e^{2i\chi t}+\nu^2\,e^{4i\chi t}}
{2 + 4\nu + \nu^2}\right) 
e^{i(4\chi t+\nu\,\sin \,2\chi t)}.
\label{z2}
\end{equation}
It is evident that these expressions are 
already considerably more involved than that in
Eq. (\ref{xt}) for $m = 0$. In particular, we note the
occurrence, in the exponents, of secular terms in $t$ apart from 
sinusoidal terms involving higher harmonics. Similar remarks apply to 
the expectation value of $p$ in these cases. The time-dependence of
the higher moments of $x$ and  $p$ are even more involved. We do not
go into these here.

In general, 
the  effects of the departure from 
coherence of the initial wave packet on its 
subsequent dynamics are not 
significant if $\nu \gg m$, since the overall exponential factor
$\exp\,[-\nu(1 - \cos \, 2\chi t)]$ 
ensures that the expectation values are essentially static except for 
sudden changes at times close to revivals and fractional revivals. 
This is not valid for smaller values of $\nu$. 
In particular, 
squeezing and higher-order squeezing effects depend crucially on the 
precise nature of the initial state, as we shall now see.
     
\section{Squeezing and Higher-Order Squeezing}

We begin by summarizing the condition for $q^{\rm th}$-power 
amplitude-squeezing 
of a quantum state and 
outlining the squeezing properties exhibited
by an initial CS as it propagates 
through a Kerr-like medium\cite{du}. We 
then examine in detail the squeezing properties exhibited by an initial 
PACS as it evolves in the medium, 
and compare its behavior with that 
of a CS.

One first defines the two quadrature variables 
\begin{equation}
Z_1= \frac{(a^q+a^{\dagger q})}{\sqrt{2}},\,\,
Z_2= \frac{(a^q-a^{\dagger q})}{i\sqrt{2}}\quad (q=1,\,2,\,3,\ldots).
\label{Z1Z2}
\end{equation}
The generalized uncertainty principle gives  
\begin{equation}
(\varDelta Z_1)^2 \,(\varDelta Z_2)^2 \geq
\textstyle{\frac{1}{4}}\,\big|\aver{\,[Z_1,Z_2]\,}\big|^2,
\label{uncertprinciple}
\end{equation}
where $\varDelta Z_i$ is the standard deviation of $Z_i\,$, and the
expectation values refer to those in the state concerned.
The state is said to be $q^{\rm th}$-power amplitude-squeezed  
in the variable $Z_1$ if
\begin{equation}
(\varDelta Z_1)^2 < \textstyle{\frac{1}{2}}\,
\big|\aver{\,[Z_1\,,\,Z_2]\,}\big|.
\label{Z1squeezing}
\end{equation}
Amplitude squeezing in $Z_2$ is similarly defined. 
We write $[a^q\,,\, a^{\dagger q}] = F_q(N)$ (this is a certain 
polynomial of order 
$(q-1)$ in the number operator $N$) and define the quantity 
\begin{equation}
D_q(t) =\frac{(\varDelta Z_1)^2-
\frac{1}{2}\aver{F_q(N)}}{\frac{1}{2}\aver{F_q(N)}},
\label{Dqone}
\end{equation}
where the time-dependence has been indicated explicitly 
to remind us that the
expectation values involved are those in the instantaneous state of
the system. It is easily seen that the state is $q^{\rm th}$-power
amplitude-squeezed in $Z_1$ if $-1 \leq D_q < 0$. 
We can rewrite Eq. (\ref{Dqone}) in terms of $a^q$ and 
$a^{\dagger q}$ as 
\begin{equation}
D_q(t) 
=\frac{2\,\left[{\rm Re}\,\aver{a^{2q}} 
-2 \,\big({\rm Re} \,\aver{a^q} \big)^2+ \aver{a^{\dagger q} \,a^{q}}\right]}
{\aver{F_q(N)}}.
\label{Dqtwo}
\end{equation} 
When the initial state is the CS 
$\ket{\alpha}$, the (time-dependent) expectation values on the RHS can be
evaluated\cite{du}  
using Eq. (\ref{csexpect}). 

We are interested, in particular, in examining 
whether fractional revivals are accompanied by 
any significant degree of squeezing and higher-order squeezing. For
this purpose, we focus on $D_q(t)$ at the instant $t=\pi/(2 
\chi) = \frac{1}{2}T_{\rm rev}\,$, 
corresponding to a 2-sub-packet fractional revival. 
After simplification, we find that $D_q$ vanishes at this instant 
for all even values of $q$, implying that 
no even-order squeezing of the
state accompanies this fractional revival. For odd values of $q$,
however, we find  
\begin{equation}
D_q(T_{\rm rev}/2) =\frac{2\nu^q}{\aver{F_q(N)}}
\big(\sin^2 q\theta - e^{-4\nu} \cos^2 q\theta\big),
\label{Dqthree}
\end{equation}
where $\theta$ is the argument of $\alpha \,(= \nu^{1/2}\, e^{i\theta})$.
\begin{figure} 
\includegraphics[width=3.3in,height=1.6in]{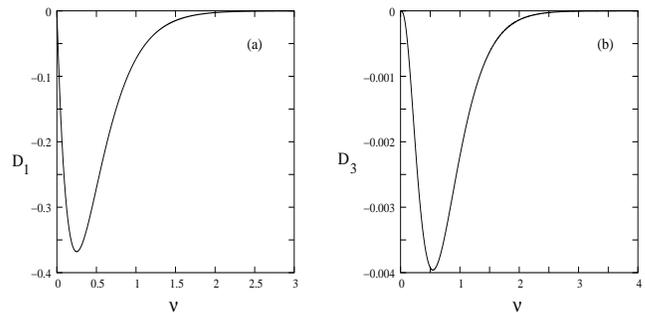}
\caption{Plots of $D_q(\frac{1}{2}T_{\rm rev})$ 
versus $\nu$ for an initial CS (with $\theta=0$), 
for (a) $q=1$ and (b) $q=3$. (Note the different ordinate scales in the
two cases.)}
\label{Dqnu}
\end{figure}
Thus squeezing (or higher-order squeezing) 
occurs at this instant provided $D_q < 0$, i.e.,  
$|\tan\,q\theta| < e^{-2 \nu}$. We illustrate this in Figs. 
\ref{Dqnu}(a) and (b), where $D_1$ and $D_3$  
are plotted as functions of $\nu$ for an initial CS with $\theta = 
0$. (We have set $\chi = 5$ in  
all the numerical results presented in this paper.) 
\begin{figure}
\includegraphics[width=3.3in,height=1.6in]{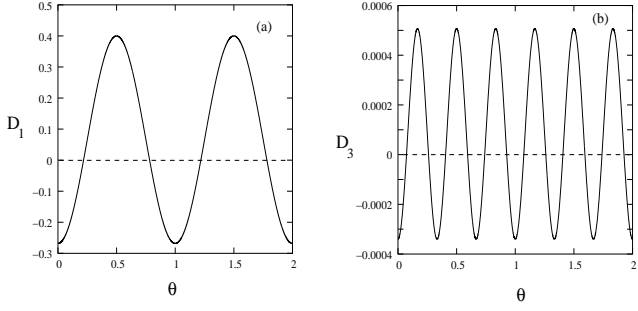}
\caption{Plots of $D_q(\frac{1}{2}T_{\rm rev})$ 
versus $\theta$ for an initial CS (with $\nu=0.1$), 
for (a) $q=1$ and (b) $q=3$.}
\label{Dqtheta}
\end{figure}
We  have also plotted $D_1$ and $D_3$ as 
functions of $\theta$ for a fixed value of $\nu\,(=0.1)$ 
in Figs. \ref{Dqtheta}(a) and (b), showing how squeezing occurs for
certain ranges of the argument of $\alpha$, when $D_q$ becomes
negative.  

We now show that if the initial state departs even marginally from 
coherence, as in a PACS with a small value of $m$, 
these results change significantly. Writing $D_q$ as  
$D_q^{(m)}$ when the expectation values in Eq. (\ref{Dqtwo}) are evaluated
for an initial state 
$\ket{\alpha , m}$, we find the following general result: setting  
$A=\frac{1}{2}L_m(-\nu)\aver{F_q(N)}$, we have 
\begin{widetext}
\begin{eqnarray}
A D_q^{(m)}(t) &=&
e^{-\nu(1-\cos4\chi qt)}\sum_{n=0}^{m}\binom{m+2q}{n+2q}
\,\frac{\nu^{n+q}}{n!} \cos\,\Big(2(2m+2n + q -1)\chi qt+\nu\sin4\chi qt
-2q\theta\Big)\nonumber\\
&-& \frac{2e^{-2\nu(1-\cos2\chi qt)}}{L_m(-\nu)}
\biggl\{\sum_{n=0}^{m}\binom{l+q}{n+q}\,\frac{\nu^{n+q}}{n!}
\cos\,\Big((q-1+2m+2n)\chi qt+\nu\sin\,2\chi qt-q\theta\Big)
\biggr\}^2\nonumber\\
&+&\sum_{n=n_{\rm min}}^{q}\binom{q}{n}\frac{m!}{(m-q+n)!}\,\nu^nL_m^n(-\nu),
\label{Dqmone}
\end{eqnarray}
\end{widetext}
where $n_{\rm min} = {\rm Max}\,(0\,,\,q-m)$. This generalizes the 
expression for $D_q$ obtained\cite{du} for a coherent state, which
corresponds to $m = 0$.

\begin{figure}
\includegraphics[width=3.3in,height=1.6in]{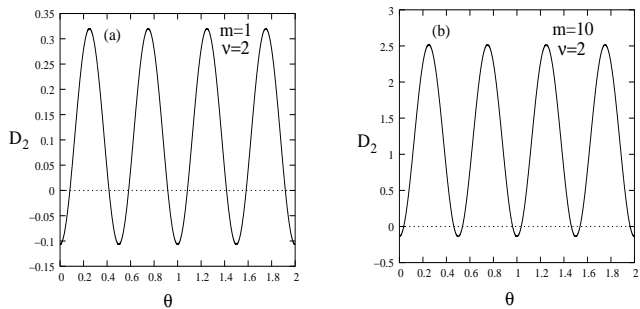}
\caption{Plots of $D_2(\frac{1}{2}T_{\rm rev})$ 
versus $\theta$ (with $\nu = 2$), 
for an initial PACS with (a) $m=1$ and (b) $m=10$.}
\label{D2mtheta}
\end{figure}

As before, we examine $D_q^{(m)}$ at $t = \pi/(2\chi)$ for the
possibility of squeezing and higher-order squeezing. The foregoing
expression reduces at this instant of time to 
\begin{eqnarray}
\lefteqn{A D_q^{(m)}(T_{\rm rev}/2) = 
(-\nu)^{q}\,L_m^{2q}(-\nu)\,\cos\,2q\theta}\nonumber\\
&-&\frac{2e^{-2\nu(1-\cos q\pi)}}{L_m(-\nu)}\nu^{q}
\Big\{L_m^q\big((-1)^q\nu\big)\Big\}^2
\cos^2 q\theta \nonumber\\
&+&\sum_{n=n_{\rm min}}^{q}\binom{q}{n}\frac{m!}{(m-q+n)!}\,\nu^nL_m^n(-\nu).
\label{Dqmtwo}
\end{eqnarray}
We use this to analyze various cases numerically.
In contrast to what happens for an initial CS, it turns out that there
is no {\it odd}-power amplitude-squeezing for an initial
PACS. Even-power amplitude-squeezing does occur, though, for
sufficiently large values of $\nu$. 
\begin{figure}
\includegraphics[width=3.3in,height=1.6in]{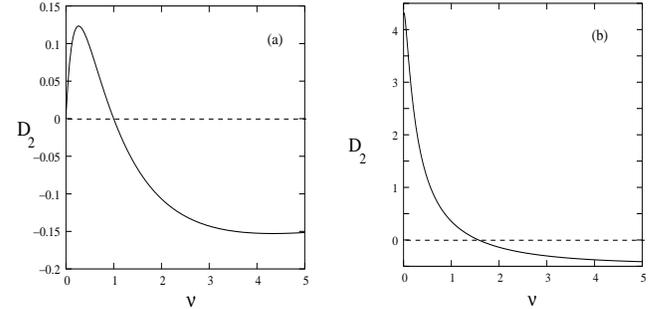}
\caption{Plots of $D_2(\frac{1}{2}T_{\rm rev})$
versus $\nu$ (with $\theta=0$), 
for an initial PACS with (a) $m=1$ and (b) $m=10$.}
\label{D2mnu}
\end{figure}
This is illustrated in
Figs. \ref{D2mnu}(a) and (b), which show the range of $\nu$ for which
$D_2$ falls below zero. Figures \ref{D2mtheta}(a) and (b) depict the
variation of $D_2$ with the phase angle $\theta$ for a fixed value of
$\nu$. 

\begin{figure}
\includegraphics[width=3.0in,height=3.0in]{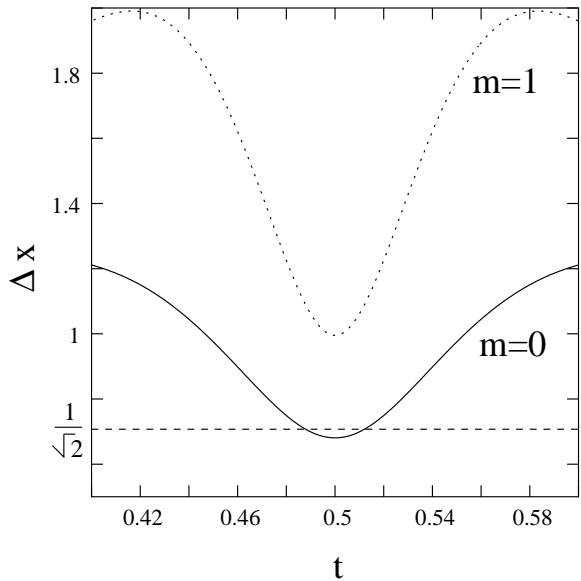}
\caption{$\varDelta x$ versus time in units of $T_{\rm rev}$, 
in the case $x_0 =
  \sqrt{2}\,,\,p_0 = 0$.}
\label{deltaxt}
\end{figure}

Turning to the extent of squeezing as a function of time,
Fig. \ref{deltaxt} compares the temporal variation of the standard deviation 
$\varDelta x$ for an initial CS ($m = 0$) and an initial PACS ($m =
1$), with $\alpha = 1$ (hence $\theta = 0$).  
The horizontal dashed line demarcates the level below which 
the state is squeezed. It is evident that squeezing in the rigorous 
sense accompanies the
fractional revival at $t = 
\frac{1}{2}T_{\rm rev}$ 
when the initial
state is a CS. This feature is suppressed when it is a PACS, although  
$\varDelta x$ does dip down considerably around this fractional
revival. 

We have focused on $q^{\rm th}$-power amplitude-squeezing at
fractional revivals, as this turns out to  provide rather more
discriminatory signatures of higher-order squeezing effects 
than the other alternative, namely, 
Hong-Mandel squeezing\cite{hong}. However, a few remarks on the
latter are in order here. 
The relevant variables in the case of Hong-Mandel squeezing
are 
$(a+a^\dagger)^q)/\sqrt{2}$ and $(a-a^\dagger)^q/(i\sqrt{2})$, i.e., 
essentially the $q^{\rm th}$ powers of  $x$ and $p$. 
For $q=1$, of course, Hong-Mandel 
squeezing is the same as amplitude-squeezing, but the two kinds of
squeezing differ for $q \geq 2$. 
\begin{figure}
\includegraphics[]{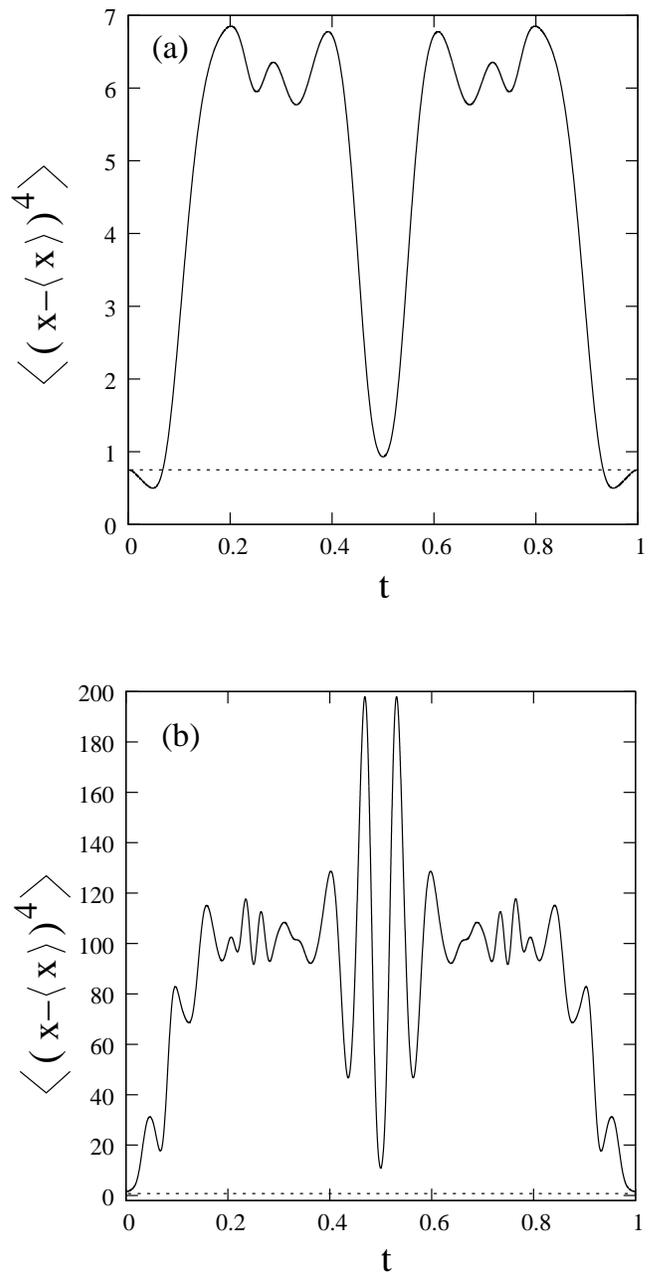}
\caption{Plots of $\aver{(x-\aver{x})^4}$ 
versus time in units of $T_{\rm rev}$ for initial states
 (a) $\ket{\alpha}$  and (b) $\ket{\alpha,5}$, 
with $\nu = 1$. Note the very different
ordinate scales in the two cases.}
\label{4thmoment}
\end{figure}
Figs. \ref{4thmoment}(a) and (b) show how 
$\aver{(x-\aver{x})^4}$,   
the fourth moment of $x$ about its mean value, varies  
over a revival period for an initial CS and PACS, respectively. 
The horizontal dotted lines indicate the bound on 
$\aver{(x-\aver{x})^4}$,   
below which fourth-order
Hong-Mandel squeezing occurs in this quadrature. 
An initial CS exhibits such squeezing near revivals, and 
comes close to doing so near the fractional revival at  
$\frac{1}{2}T_{\rm rev}$, but does not actually do so. 
An initial PACS does not display such higher-order squeezing at any
time, although 
$\aver{(x-\aver{x})^4}$ attains its lowest value at
revival times. However, fractional revivals are marked by 
rapid oscillations of $\aver{(x-\aver{x})^4}$,  
these being most pronounced around the $2$-sub-packet fractional
revival. These features are enhanced further in the case of 
initial states with larger values of
$m$. 

\section{The Wigner function and the non-classicality indicator}

\begin{figure}
\includegraphics{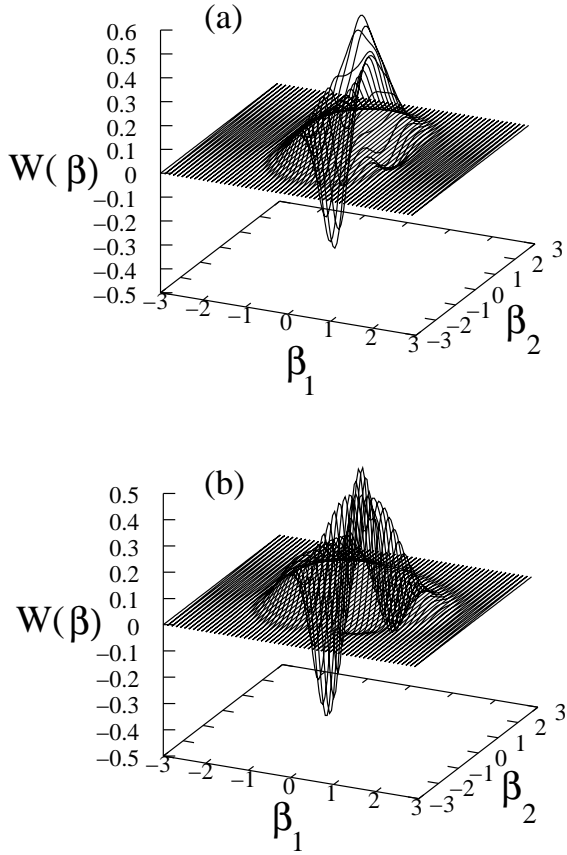}
\caption{Plots of the Wigner function corresponding to an initial state 
$\ket{\alpha}$ with $\alpha = 1$, at (a) $t=\frac{1}{2}T_{\rm rev}$ and (b) 
$t=\frac{1}{3}T_{\rm rev}$. Here, and in the succeeding figures, 
$\beta_1 = {\rm Re}\,\beta\,,\,
\beta_2 = {\rm Im}\,\beta$.} 
\label{wignercs}
\end{figure}
\begin{figure}
\includegraphics{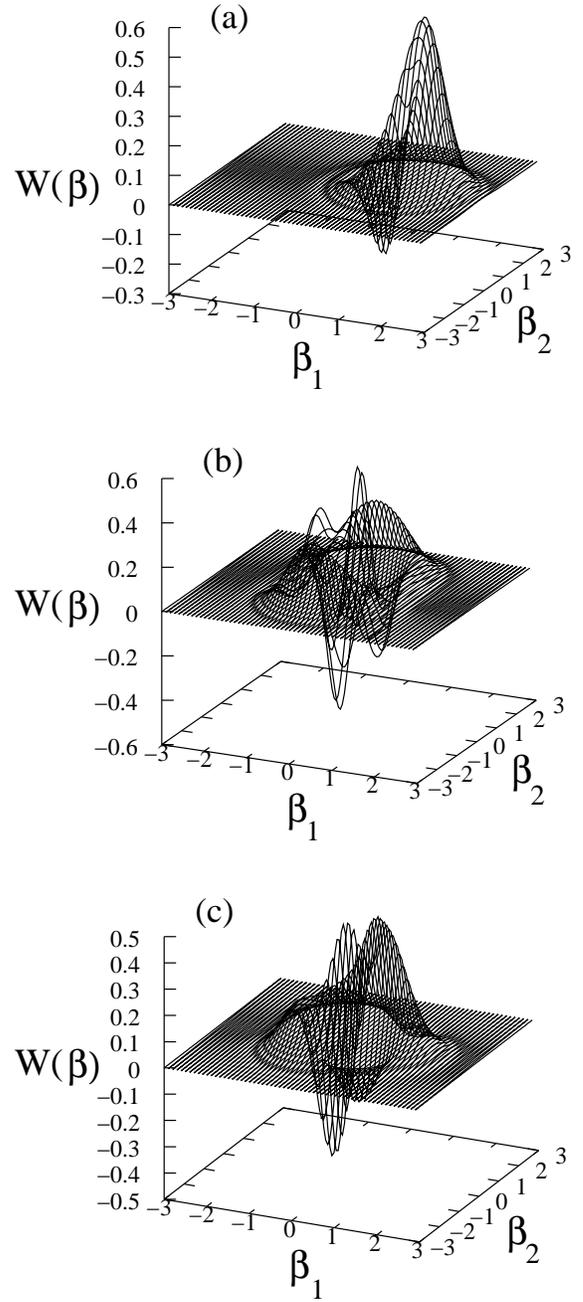}
\caption{Plots of the Wigner function corresponding to an initial state
$\ket{\alpha,1}$ with $\alpha = 1$, 
at (a) $t=0$,\, (b) $t=\frac{1}{2}T_{\rm rev}$ 
and \,(c) $t=\frac{1}{3}T_{\rm rev}$. }
\label{wigner1}
\end{figure}
Finally, we examine the 
Wigner functions corresponding to the wave packets 
at instants of fractional revivals, to quantify  
non-classical behavior during their time evolution. It is
well known that the ``extent'' to which the Wigner function becomes
negative (as a function of its complex argument) is an indicator of
the non-classicality of the state concerned. 
The normalized Wigner function $W(\beta\,;\,t)$ 
(where $\beta \in \mathbb{C}$)  
is given by\cite{brune} 
\begin{eqnarray}
W(\beta\,;\,t)
&=&\frac{2}{\pi}e^{-2|\beta|^2}{\rm Re}\biggl\{\,\sum_{l,n 
=m \atop n \geq l}^\infty 
(-1)^l(2-\delta_{l\,n})\,(l!/n!)^{1/2}\nonumber\\
&&\times(2\beta)^{n-l}\,
\rho_{l n}(t)\,
L_l^{n-l}\big(4|\beta|^2\big)\biggr\},
\label{wignerfn}
\end{eqnarray}
where $\rho_{l n}(t)$ is the density matrix element corresponding to
the state at time $t$ in the oscillator Fock basis. For an initial 
coherent state  $\ket{\alpha}$ we have the standard result
\begin{equation}
\rho_{ln}(0) =
\frac{\alpha^{*n}\,\alpha^l}
{\sqrt{l!\, n!}}\,e^{-|\alpha|^2}, 
\label{rhocs}
\end{equation}
leading to the well-known expression
\begin{equation}
W(\beta\,;\,0)=\frac{2}{\pi}e^{-2|\alpha-\beta|^2}.
\label{wignerfncs}
\end{equation}
for the Wigner function. 
This is positive definite everywhere in the complex $\beta$-plane,
justifying the appellation ``classical'' for an oscillator coherent
state $\ket{\alpha}$. The time evolution of the 
Wigner function under the Hamiltonian 
of Eq. (\ref{hamil}) may be computed readily using the representation
$\rho (t) =\sum_{l,n=0}^{\infty}\rho_{ln}(t) \ket{l}\bra{n}$ for the
density matrix. Figures \ref{wignercs}(a) and (b) show how the 
Wigner function for this initial CS 
behaves at the 
instants of the 2-sub-packet 
and 3-sub-packet fractional revivals,
respectively. 
\begin{figure}
\includegraphics{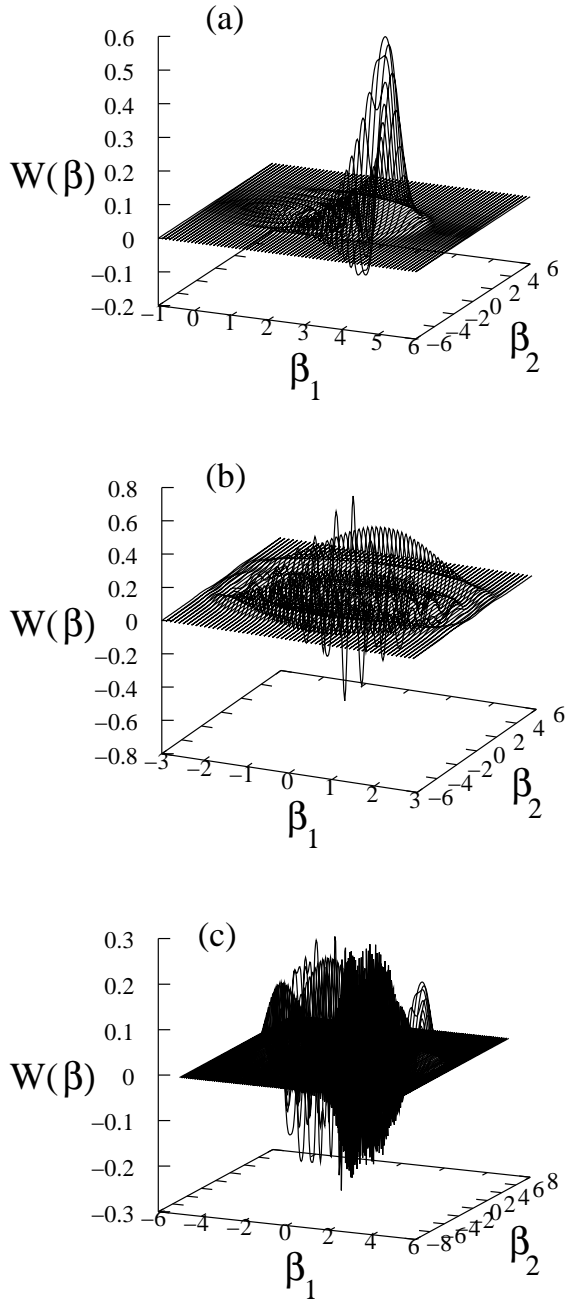}
\caption{Plots of the Wigner function corresponding to an initial state
$\ket{\alpha,10}$  at (a) $t=0$,\, (b) $t=\frac{1}{2}T_{\rm rev}$ 
and (c) $t=\frac{1}{3}T_{\rm rev}$.}
\label{wigner10}
\end{figure}

For the initial photon-added coherent state
$\ket{\alpha,m}$, one finds\cite{agar}
\begin{equation}
\rho_{l n}(0) =\frac{e^{- \nu}}{m!\,L_m(- \nu)}
\frac{\alpha^{l-m}\,{\alpha^*}^{n-m}\sqrt{l!\, n!}}{(l-m)!\,(n-m)!}.
\label{rhopacs}
\end{equation}
Correspondingly, the Wigner function at $t = 0$  
can be expressed in the closed form 
\begin{equation}
W(\beta\,;\,0)=\frac{2\,(-1)^m}{\pi L_m(- \nu)}\,
L_m\big(|2\beta-\alpha|^2\big)\,
e^{-2|\alpha-\beta|^2}.
\label{wignerfnpacs}
\end{equation}
It is to be noted that this is no longer positive definite for  all
complex $\beta$, reflecting the fact that this initial state 
is no longer
``totally'' classical, as it departs
from perfect coherence by the photons that have been ``added'' to 
$\ket{\alpha}$ to produce the PACS. 
Figures \ref{wigner1}(a), (b) 
and (c) are plots of $W(\beta\,;\,t)$ for the case 
$m=1$ at $t=0,\,\frac{1}{2}T_{\rm
  rev}$ and $\frac{1}{3}T_{\rm rev}\,$, respectively.  
The  corresponding plots for the case $m=10$ 
are shown in Figs. \ref{wigner10}(a), (b) and (c). With increasing
$m$, the oscillations of
the Wigner function in the $\beta$-plane between positive and negative
values become more pronounced, and the region of 
non-classicality becomes more extensive.

\begin{figure}
\vspace{.4cm}
\includegraphics{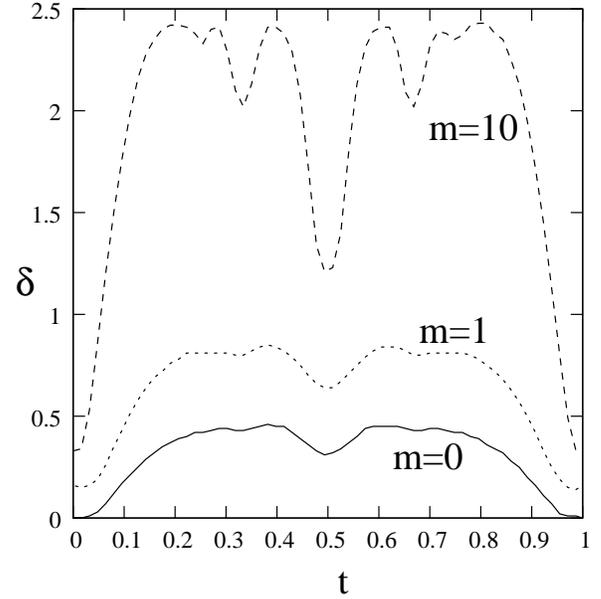}
\caption{Plot of $\delta$ versus time in units of $T_{\rm rev}$.}
\label{wignerdelta}
\end{figure}
To get an idea of the degree of non-classicality as a continuously
varying function of time
for each of the different initial states we have considered, 
it is instructive to 
consider the non-negative quantity $\delta$ defined as\cite{ken}  
\begin{eqnarray}
\delta (t) &=&\int \!d^{\,2}\beta\,
\Big(\big|W(\beta\,;\,t)\big|
-W(\beta\,;\,t)\Big)\nonumber\\
&=&\int \!d^{\,2}\beta\,\big|W(\beta_1\,;\,t)\big|-1.
\end{eqnarray}
The larger the value of $\delta$, the greater is the extent of
non-classicality of the state concerned, although of course 
$\delta$ alone does not give a complete picture of the oscillations of
the Wigner function. In Fig. \ref{wignerdelta} 
we have plotted 
$\delta$ versus $t$ for initial states $\ket{\alpha},\,\ket{\alpha,1}$, 
and $\ket{\alpha,10}$ where $\alpha$ has been set equal to unity. 
It is clear that in the interval 
between $t=0$ and $t = T_{\rm rev}\,$,  $\delta$ is least at 
the $2$-sub-packet 
fractional revival, followed by its values at the $3$-sub-packet 
and $4$-sub-packet fractional revivals. This 
feature becomes increasingly 
prominent for larger values of $m$, showing that 
the extent of 
non-classicality also increases with $m$, as expected.

This work was supported in part by the
Department of Science and Technology, India, under Project No.
SP/S2/K-14/2000.

\end{document}